\documentclass[conference]{IEEEtran}
\IEEEoverridecommandlockouts
\usepackage{cite}
\usepackage{amsmath,amssymb,amsfonts}
\usepackage{algorithmic}
\usepackage{graphicx}
\usepackage{textcomp}
\usepackage{xcolor}
\def\BibTeX{{\rm B\kern-.05em{\sc i\kern-.025em b}\kern-.08em
    T\kern-.1667em\lower.7ex\hbox{E}\kern-.125emX}}
    \usepackage[acronym, nowarn]{glossaries}
\makeglossaries

\newacronym{ble}{\textsc{BLE}}{Bluetooth Low Energy}
\newacronym{btmesh}{\textsc{BT Mesh}}{Bluetooth Mesh}
\newacronym{btsig}{\textsc{SIG}}{Special Interest Group}
\newacronym{iot}{\textsc{IoT}}{Internet of Things}
\newacronym{bemergency}{\textsc{Bluemergency}}{Bluetooth Mesh emergency network}
\newacronym{d2d}{\textsc{D2D}}{device-to-device}

\usepackage{todonotes}
\usepackage{tabularx}
\usepackage[flushleft]{threeparttable}
\usepackage{array,booktabs,ragged2e}

\usepackage{tikz}
\usetikzlibrary{calc, positioning, arrows, shapes, shadows, trees, arrows.meta, 
decorations.pathreplacing, hobby, backgrounds} 
\usepackage{adjustbox}
\usepackage{epstopdf}
\usepackage{multirow}
\usepackage{wasysym}
\usepackage{gensymb}
\usepackage{amssymb}
\usepackage{pifont}

\usepackage{dblfloatfix}
\usepackage{subcaption}
\usepackage{amsthm}

\usepackage[hyphens]{url}
\usepackage{hyperref}
\hypersetup{
    colorlinks = false,
    allcolors = {red}
}

\theoremstyle{definition}

\pgfdeclaredecoration{dashsoliddouble}{initial}{
  \state{initial}[width=\pgfdecoratedinputsegmentlength]{
    \pgfmathsetlengthmacro\lw{.3pt+.5\pgflinewidth}
    \begin{pgfscope}
      \pgfpathmoveto{\pgfpoint{0pt}{\lw}}%
      \pgfpathlineto{\pgfpoint{\pgfdecoratedinputsegmentlength}{\lw}}%
      \pgfmathtruncatemacro\dashnum{%
        round((\pgfdecoratedinputsegmentlength-3pt)/6pt)
      }
      \pgfmathsetmacro\dashscale{%
        \pgfdecoratedinputsegmentlength/(\dashnum*6pt + 3pt)
      }
      \pgfmathsetlengthmacro\dashunit{3pt*\dashscale}
      \pgfsetdash{{\dashunit}{\dashunit}}{0pt}
      \pgfusepath{stroke}
      \pgfsetdash{}{0pt}
      \pgfpathmoveto{\pgfpoint{0pt}{-\lw}}%
      \pgfpathlineto{\pgfpoint{\pgfdecoratedinputsegmentlength}{-\lw}}%     
      \pgfusepath{stroke}
    \end{pgfscope}
  }
}  

\tikzset{
  -|-/.style={
    to path={
      (\tikztostart) -| ($(\tikztostart)!#1!(\tikztotarget)$) |- (\tikztotarget)
      \tikztonodes
    }
  },
  -|-/.default=0.5,
  |-|/.style={
    to path={
      (\tikztostart) |- ($(\tikztostart)!#1!(\tikztotarget)$) -| (\tikztotarget)
      \tikztonodes
    }
  },
  |-|/.default=0.5,
}

\newcommand\doublerule{\hline\hline}
\newcommand\doubleheavyrule{\hline\specialrule{\heavyrulewidth}{\doublerulesep}{0.0em}}

\newcommand\copyrighttext{%
  \footnotesize \copyright{} 2019 IEEE. Personal use of this material is permitted. Permission from IEEE must be obtained for all other uses, in any current or future media, including reprinting/republishing this material for advertising or promotional purposes, creating new collective works, for resale or redistribution to servers or lists, or reuse of any copyrighted component of this work in other works.}
  
\newcommand\copyrightnotice{%
\begin{tikzpicture}[remember picture,overlay]
\node[anchor=south,yshift=10pt] at (current page.south) {\fbox{\parbox{\dimexpr\textwidth-\fboxsep-\fboxrule\relax}{\copyrighttext}}};
\end{tikzpicture}%
}

\begin{document}

\title{Bluemergency: Mediating Post-disaster Communication Systems using the Internet of Things and Bluetooth Mesh\\
}

\author{
	\IEEEauthorblockN{
        Flor~\'{A}lvarez, %\IEEEauthorrefmark{1}, 
        Lars Almon, %\IEEEauthorrefmark{1}, 
        Hauke Radtki and %\IEEEauthorrefmark{1} and 
        Matthias Hollick%\IEEEauthorrefmark{1} 
	}
	\IEEEauthorblockA{
		Secure Mobile Networking Lab, TU Darmstadt, Germany\\
		Email: \{falvarez, lalmon, hradtki, mhollick\} @seemoo.tu-darmstadt.de\\
	}
}

\maketitle
\copyrightnotice

%!TEX root = ../ghtc_btmesh_2019.tex
\begin{abstract}
Mobile devices have shown to be very useful during and post disaster. 
If the communication infrastructure breaks down, however, they become almost useless as most services rely on Internet connectivity. 
Building post-disaster networks based purely on smartphones remains a challenging task, and, as of today, no practical solutions exist. 
The rapidly growing \gls{iot} offers the possibility to improve this situation. 
With an increase in smart spaces such as smart homes and smart offices, we move towards digital cities that are deeply penetrated by \gls{iot} technology. 
Many \gls{iot} devices are battery powered and can aid in mediating an emergency network. 
In scenarios where the electrical grid is still operational, yet communication infrastructure failed, non-battery powered \gls{iot} devices can similarly help to relief congestion or build a backup network in case of cyber attacks. 
With the recent release of the Bluetooth Mesh standard, a common interface between mobile devices and the \gls{iot} has become available. 
The key idea behind this standard is to allow existing and new devices to build large-scale multi-hop sensor networks. 
By enabling hundreds of devices to communicate with each other, \gls{btmesh} becomes a practical technical solution for enabling communication post disaster. 
In this paper, we propose a novel emergency network concept that utilizes the parts of digital cities that remains operational in case of disaster, thus mediating large-scale post-disaster device-to-device communication. 
Since the Bluetooth Mesh standard is backwards compatible to Bluetooth 4.0, most of today's mobile devices can join such a network. 
No special hardware or software modifications are necessary, especially no jail-breaking of the smartphones. 
\end{abstract}


%!TEX root = ../ghtc_btmesh_2019.tex
\section{Introduction}
Usage of the \gls{iot} has grown rapidly in recent years \cite{gazis2017survey, hui2017major}.
It is estimated that by 2025 the installed base of \gls{iot} connected devices will grow to almost 75 billion sensing devices.
In fact, the \gls{iot} concept covers a wide range of solutions\cite{statistaIoT}.
Smart offices \cite{khajenasiri2017review} and smart homes \cite{malche2017internet} represent a prominent \gls{iot} use case. On the one hand, smart office solutions aim to provide a more comfortable and energy efficient workspace, where sensors, e.g., allow to adjust the light or heat according to the current measurement of an office \cite{mikulecky2016workshop, murthy2018smart}.
On the other hand, smart home systems integrate and connect common home devices such as lighting, 
heating, a refrigerator, etc., to offer an automated environment, in which many house features can be 
controlled and monitored locally as well as remotely \cite{malche2017internet}.
However, these smart environments mainly require the Internet to enable the communication and interaction between the smart objects.

In the last decade, Bluetooth and especially \gls{ble} have risen to become one of the most used communication technologies for the \gls{iot} \cite{Collotta2018bt5}. 
On July 19, 2017, the Bluetooth \gls{btsig} presented \gls{btmesh} \cite{MeshProfile2017, MeshModel2017}: a protocol that allows devices to communicate in a mesh based network topology.
The key idea behind this standard is to allow existing and new devices to build large-scale multi-hop sensor networks.
In addition, the standard also provides a backward compatibility, i.e., mobile devices compatible with Bluetooth 4.0 or later may also send messages in a \gls{btmesh} network.

\begin{figure*}
\vspace{+0.1cm}
	\centering
	\includegraphics[width=0.9 \textwidth]{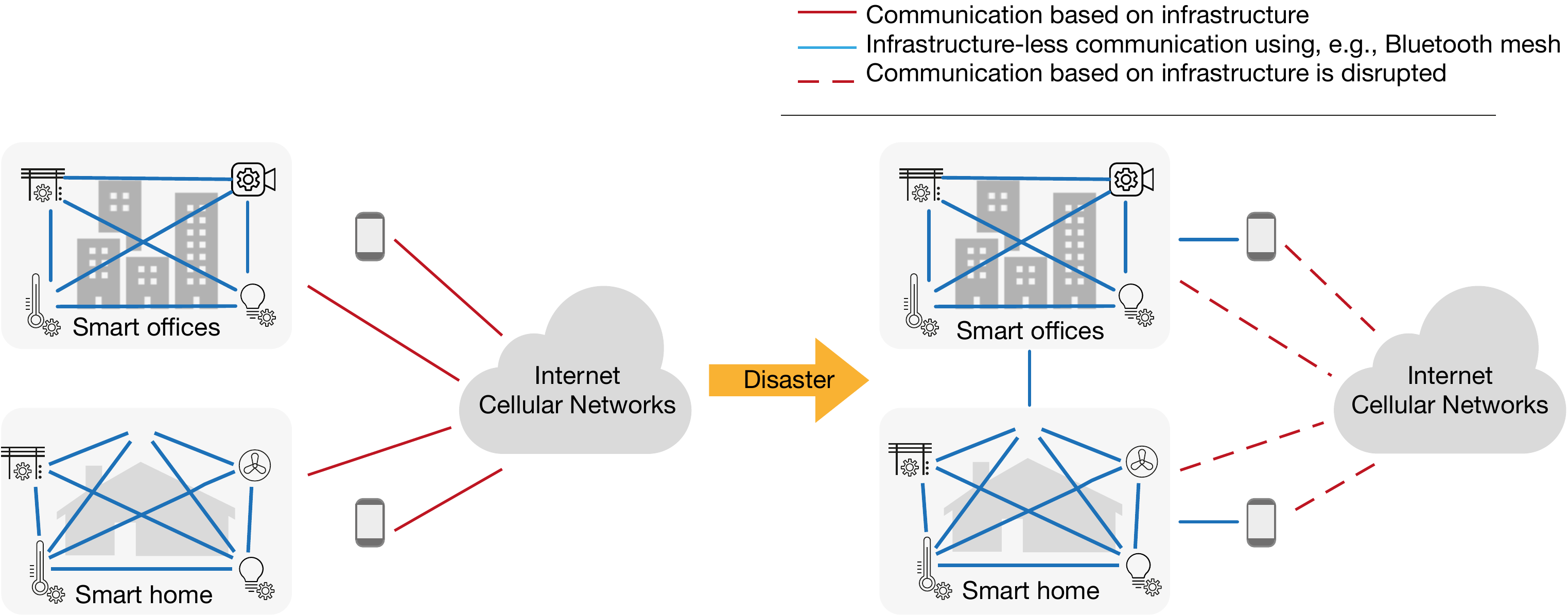}
	\caption{Integration of IoT solutions into post-disaster systems.}
	\label{iotdisaster}
\end{figure*}

By enabling hundreds of devices to communicate with each other, \gls{btmesh} becomes a practical technical solution for enabling communication post disaster.
In fact, the integration of mobile devices into these mesh networks opens up new possibilities for building post disaster communication systems as depicted in Fig.~\ref{iotdisaster}.

First, since \gls{btmesh} allows many-to-many communications, there is not a single point of failure.
Second, the mesh devices are typically sensors with an integrated power source (e.g., battery), i.e., most of them remain functional even during a blackout or if the electrical grid is severely impaired.  
Third, the backward compatibility facilities the connection of existing Bluetooth devices to an existing mesh network without the need of additional hardware or significant software changes.
Finally, by including mobile devices it is possible to build self-organizing distributed wireless networks by leveraging the parts of digital cities that remain operational, thus enabling the population to communicate without relying on a centralized infrastructure.

This work proposes the \gls{bemergency}, a practical solution to mediate device-to-device communication in post disaster scenarios by harnessing the IoT devices that remain operational in case of disaster.

The main contributions of this paper are as follows.
\begin{itemize}
    \item We provide an overview of the new \gls{btmesh} standard, highlighting key features relevant for post-disaster systems.
    \item We introduce the \gls{bemergency} concept, a practical solution to allow forming emergency networks based on mobile devices and \gls{btmesh} devices. 
    As part of \gls{bemergency}, we propose a \gls{btmesh} vendor model for allowing the data exchange between mobile devices using the mesh network.
    \item We implement an Android application to demonstrate and test the feasibility of \gls{bemergency} in practice.
    \item Finally, we evaluate the performance of our solution in two smart environments: a smart office and a smart home.
\end{itemize}

This paper is structured as follows. First, we summarize related work in Section \ref{relatedwork}. 
In Section \ref{background} we briefly introduce the new Bluetooth standard and its terminology.
In Section \ref{usecase} we detail our \gls{btmesh} emergency concept.
Section \ref{system} describes our proof-of-concept implementation.
The results of the experimental evaluation are presented in Section \ref{evaluation}.
Finally, Section \ref{conclusion} concludes this work, discussing several points for future work.
%!TEX root = ../ghtc_btmesh_2019.tex
\section{Related work}
\label{relatedwork}

So far, existing work in the field of \gls{btmesh} focuses mainly on the performance evaluation of such a network in smart environments, e.g., for building automation applications \cite{martinez2018smart}, proposing a smart-home architecture to demonstrate the feasibility of using this standard in smart home control systems \cite{wan2018smart}, or for smart cities \cite{veiga2019proposal}, etc. 
This paper aims at providing a solution to build self-organizing emergency networks without relying on a central infrastructure.

The importance of self-organizing mobile ad-hoc networks after a disaster has been widely 
studied in recent researches. 
There are already a quite a number of studies focusing on post-disaster systems based on self-organizing mobile ad-hoc networks \cite{gardner2011serval,gardner2012meshms,verma2015manet,lieser2017architecture,alvarez2018conducting}. 
These solutions leverage mobile ad-hoc networks (MANET) or delay-tolerant networks (DTN) technology to facilitate message routing/forwarding/spreading in the affected area.
However, most of them either require the installation of additional hardware or software modifications are necessary, e.g., jail-breaking off-the-shelf devices, to enable mobile devices to be part of a wireless mesh networks.

In contrast, this work proposes a practical solution based on the \gls{btmesh} standard to facilitate a device-to-device communication in post-disaster scenarios. The proposed solution involves devices that typically remains functional 
after a disaster, i.e., by utilizing the infrastructure from smart environments. 
We present an experimental evaluation of such a system using well-known \gls{iot} application scenarios, namely, smart office and smart home. Our proof-of-concept considers heterogeneous devices, including devices that support the \gls{btmesh} stack, and devices which 
can communicate with the network without the need to implement the whole stack. 
%!TEX root = ../ghtc_btmesh_2019.tex
\section{Background}
\label{background}
This section briefly introduces the key features and capabilities of \gls{btmesh} technology and details the underlying concept.
\subsection{Concept}
\label{concept}
\gls{btmesh} is a flooding-based network that uses the publish/subscribe model for the data exchange, i.e., devices can send (\textit{publish}) and/or receive (\textit{subscribe}) certain information according 
to their interests.
These networks can support up to 32767 devices, and a maximum of 128 hops are possible. 
An unsegmented message has a maximal size of 29 bytes, with the maximum application data payload size being 11 bytes. %11 from these bytes represent the maximal application data payload size.
The standard includes two different bearers: 
\textit{(i) advertising bearer:} is a non-connectable advertisement bearer which uses a new type of \gls{ble} advertisement packet to communicate, and 
\textit{(ii) GATT (\textit{Generic Attributes)} bearer: } is a connection-oriented bearer, that provides backwards compatibility, i.e., it allows any Bluetooth device compatible with GATT to also be part of a mesh network. 
This bearer utilizes the Proxy Protocol \cite{MeshProfile2017} to exchange data between two devices using a GATT connection.

\voffset+0.1cm
\subsubsection{Network Elements}
\label{nodefeature}
\begin{figure}[!b]
	\centering
	\includegraphics[width=0.5 \textwidth]{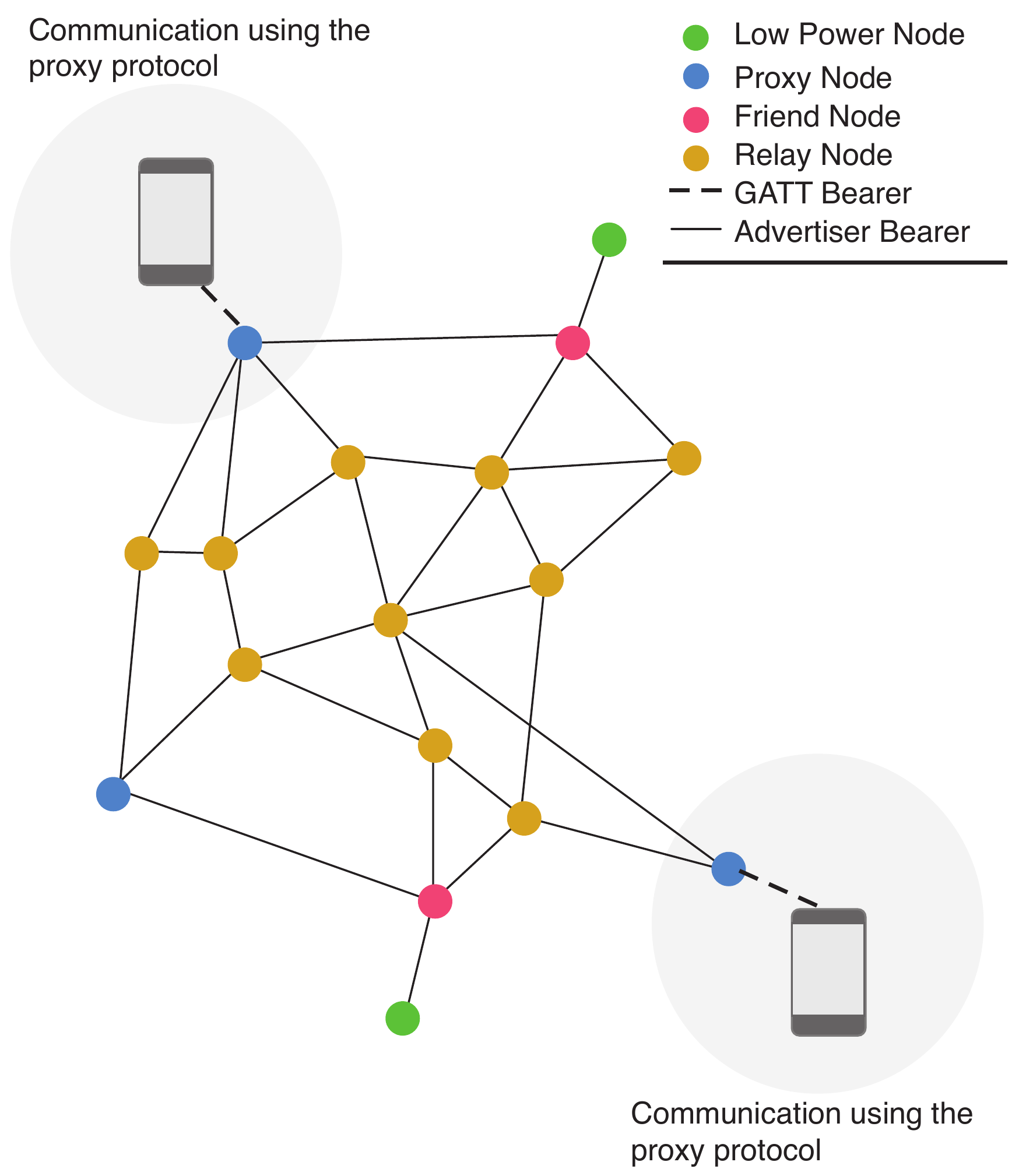}
	\caption{Bluetooth Mesh concept.}
	\label{btmesh}
\end{figure}

In order to build a \gls{btmesh} network, the devices need to be provisioned. 
During the provisioning process a device---known as a \textbf{\textit{provisioner}}---distributes necessary security material to an unprovisioned device that wants to join the network. 
A provisioned device---also called a \textbf{\textit{node}}---can send and receive mesh messages. 
Mesh nodes can support one or more additional features: 
\begin{itemize}
    \item \textbf{Relay nodes:} can also retransmit received mesh messages using the advertising bearer.
    \item \textbf{Proxy nodes:} can communicate using both communication bearers: GATT and Advertising.
    \item \textbf{Low Power nodes:} are power limited nodes that scan the communication channel at a reduced duty cycles.
    \item \textbf{Friend nodes:} stores messages addressed to Low Power nodes and retransmits them to those nodes later. 
\end{itemize}
Fig.~\ref{btmesh} shows a possible \gls{btmesh} network configuration with several nodes and all features supported by a mesh node.
For communication these nodes can either use advertising or GATT bearer.
Additionally, mobile devices that do not support \gls{btmesh} can communicate with the network using an additional communication protocol---known as \textbf{\textit{proxy protocol}}---specified in \cite{MeshProfile2017}.
	%\vspace{-0.1cm}
%	\vspace{-0.1cm}
\subsubsection{Models}
The basic functionality of nodes is defined by multiple services. 
Services---also called \textbf{\textit{models}}---can be generic or vendor specific. 
A model is identified by 16-bit (generic) or 32-bit ID (vendor specific). 
The generic models are specified in the standard. 
A common example is the generic OnOff model, where a state can be set to on or off. 
On the other hand, vendor models can be designed and implemented freely. 
In most cases, generic and vendor models are implemented using the client/server concept: a server model provides a service and a client model consumes this service.

\vspace{+0.1cm}
\vspace{+0.1cm}
\subsubsection{Security}
The \gls{btmesh} specification also considers security as mandatory, so all messages exchanged between devices on the network must be encrypted. 
The standard defines two keys used to secure messages, namely, network keys \textbf{\textit{NetKey}} and application keys \textbf{\textit{AppKey}}.
The \textbf{\textit{NetKey}} allows devices to participate in one or more subnets, as well as in different mesh networks. 
The \textbf{\textit{AppKey}} enable devices to receive or to send messages related to a given application domain.
Regarding to privacy, the standard recommends the implementation of network PDU obfuscation in order to prevent tracking of nodes in a mesh network.

\subsubsection{Backward Compatibility}
Bluetooth devices compatible with Bluetooth 4.0 or later, which do not implement the Bluetooth Mesh stack, can communicate with nodes from a \gls{btmesh} network using a GATT connection. 
To this end, these devices need to implement the proxy protocol. 
This protocol defines two node roles: server and client.
The proxy server is a node supporting both bearers, and a proxy client node supports only the GATT bearer.
For example, mobile devices act as proxy clients to transmit and receive mesh network packets over the connection-oriented GATT bearer. 
In addition, a mesh node that supports the proxy feature can act as a proxy server, and relay mesh network packets from a proxy client to other nodes in the network. 

%!TEX root = ../ghtc_btmesh_2019.tex
\section{Bluetooth mesh emergency network}
\label{usecase}
In this section, we introduce our post-disaster solution that includes 
devices from \gls{iot} solutions such as smart offices and smart homes to build emergency 
networks.

\subsection{Concept}
Natural or man-made disasters can occur at any time. A typical problem in the aftermath of a disaster is the damage of infrastructure, where mainly information and communication systems 
are affected and partially or totally unavailable. 
As a result, millions of people in need for help are isolated, especially during the crucial first hours. This disruption of 
communication also hinders the coordination of the relief efforts. 

But, if we consider \gls{iot} devices from smart offices and smart homes, we can build an emergency network to allow a device-to-device communication.
Typically, these end devices are constrained sensors with a 
integrated power source (e.g., battery), which allow them to be available even if a central power infrastructure is knock out.

\begin{figure*}[t]
\vspace{+0.1cm}
%	\centering
\hspace*{\fill}%
	\begin{subfigure}{0.29\linewidth}
		\includegraphics[width=\textwidth]{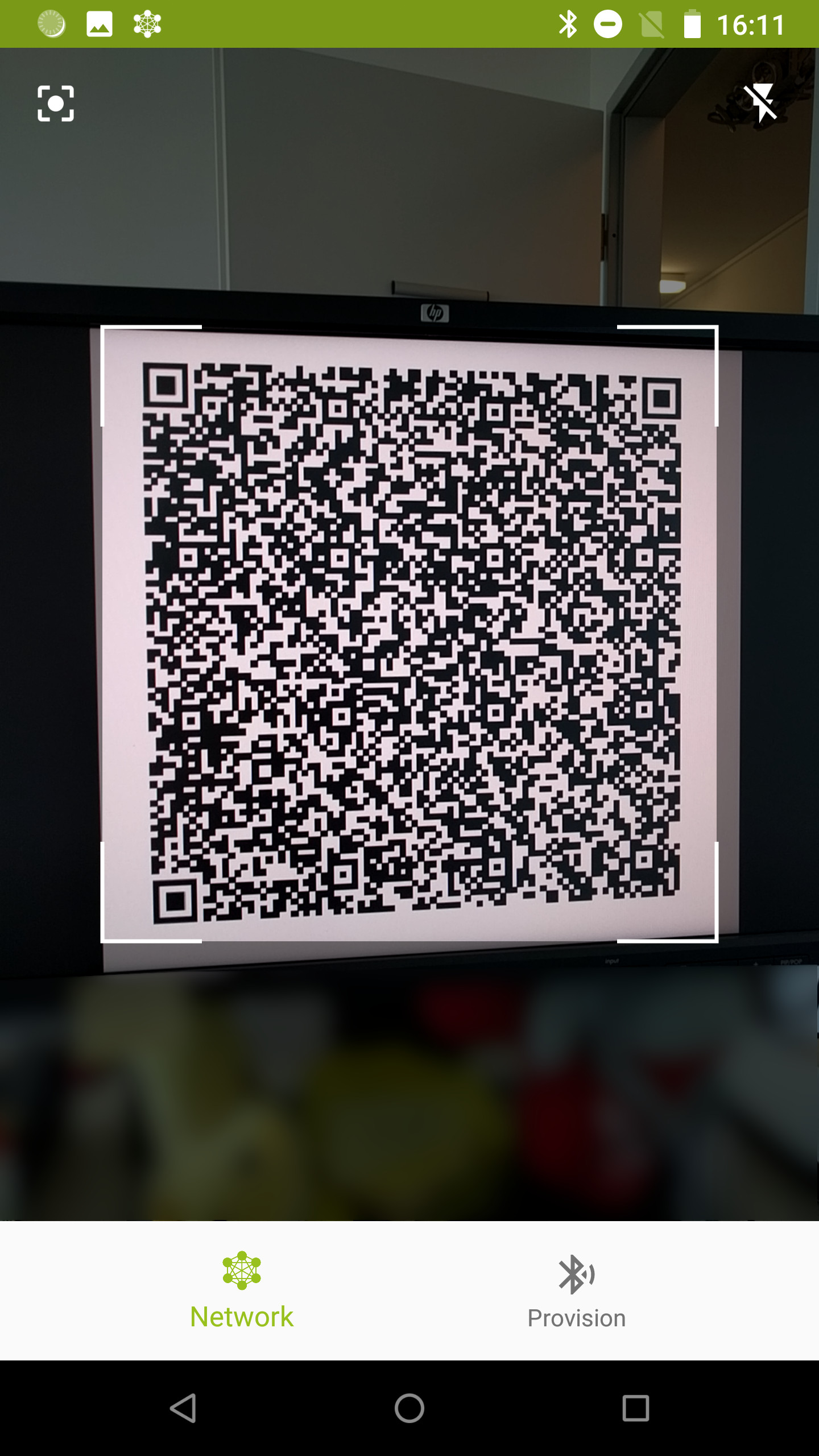}
		\caption{Joining the network via QR-Code.}
	\end{subfigure}
	\begin{subfigure}{0.29\linewidth}
		\includegraphics[width=\textwidth]{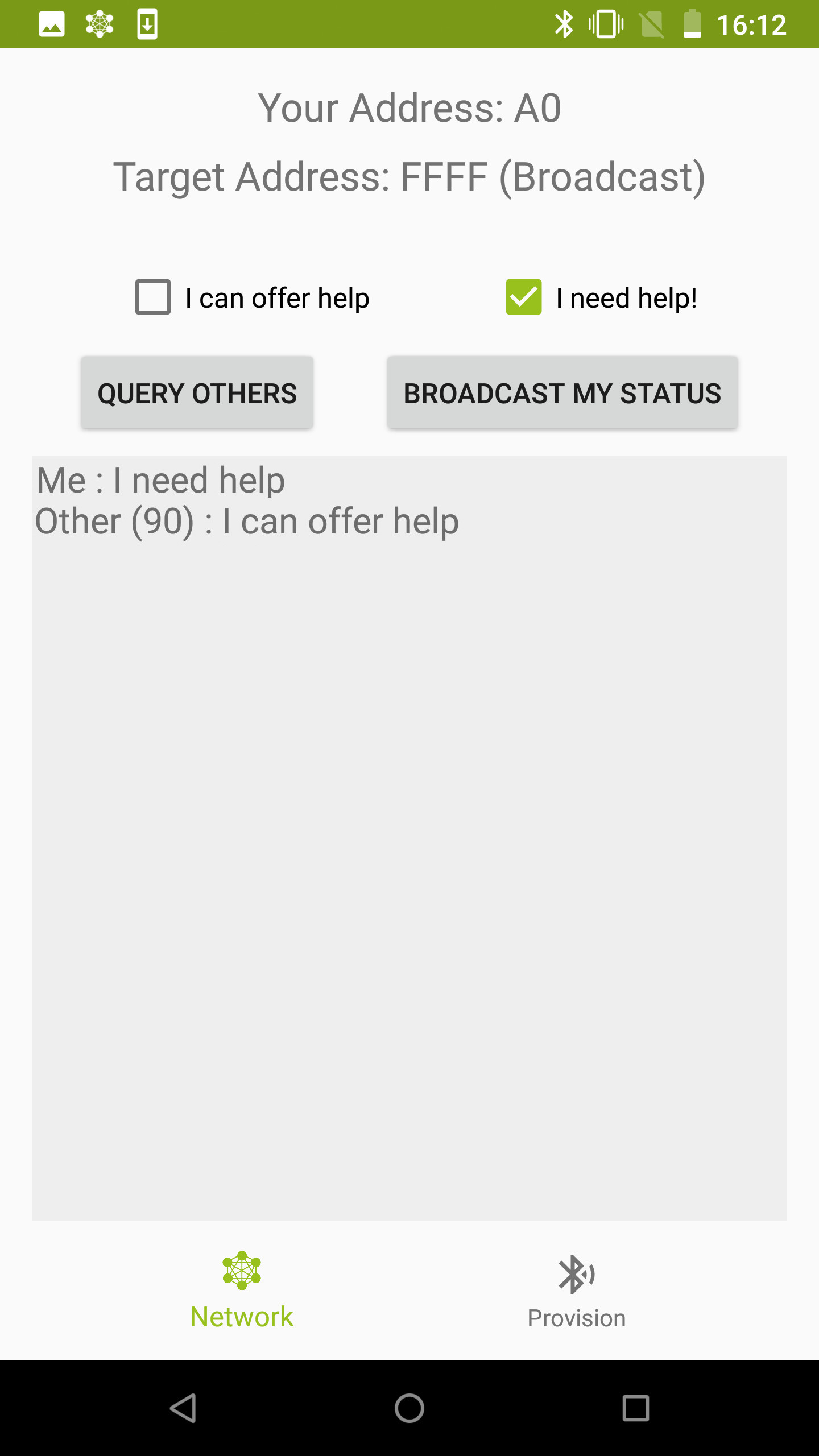}
		\caption{Requesting Help.}
	\end{subfigure}
	\begin{subfigure}{0.29\linewidth}
		\includegraphics[width=\textwidth]{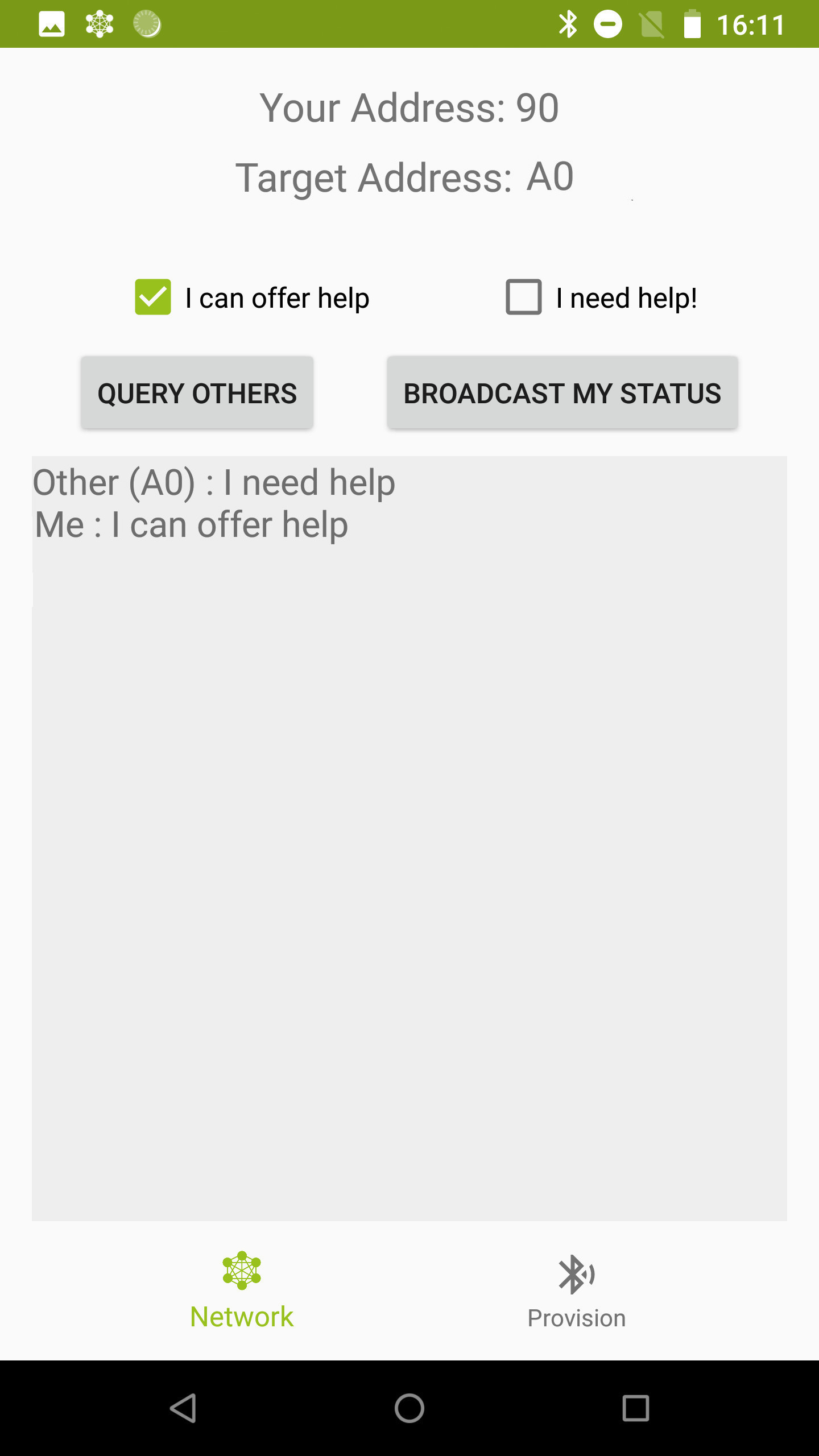}
		\caption{Offering Help.}
	\end{subfigure}
\hspace*{\fill}%
	\caption{Screenshots of our Android application developed as proof-of-concept.}
	\label{screenshots}
\end{figure*}

\subsection{Relevant features}
\gls{bemergency} is designed to complement existing self-organizing network solutions. By utilizing the \gls{btmesh} networks, our solution fulfills the most representative requirements for emergency networks \cite{miranda2016survey}.
In general, we satisfy the following requirements:
\begin{enumerate}
\item \textit{Resilience:} An important requirement for self-organizing emergency networks is the capability to provide an acceptable level of communication to cope in absence of infrastructure. A system based on a mesh topology offers resilience, as there is not a single point of failure. In contrast, each device is able to communicate with other devices and also relays messages.
\item \textit{Basis emergency services: } After a disaster, the communication needs focus mainly on the exchange of small but vital data, such as help messages or telling family and friends that you are safe. By implement a \gls{btmesh} vendor model, we can support services commonly used in emergency situations \cite{lieser2017architecture}. 
\item \textit{Self-organized: } The self-organized capability of \gls{btmesh} allows to build a system easily adaptable and relocatable which improves the reliability of a \gls{btmesh} based emergency network.
\item \textit{Mobility:} The integration of mobile devices in \gls{btmesh} smart environments facilities the creation of networks with a variable topology.
\item \textit{Interoperability: } One of the main limitations of existing emergency network is the missing interoperability between the different implementations because of the lack of a common standard. In contrast, \gls{bemergency} resolves this issue by proposing a solution based on a standard.
\end{enumerate}
\subsection{Services} 
We propose a \gls{btmesh} vendor model to facilitate the data exchange between mobile devices in the 
emergency network. Currently, we provide only two services commonly used in emergency situations \cite{lieser2017architecture}, namely: \textit{SOS Emergency Messages}, and \textit{I am Alive Notifications}. 
Table \ref{tab:vendormodel} summarizes the data structure of each packet using our model.

\begin{table}[h]
\renewcommand{\arraystretch}{1.5}
	\caption{Data structure for the emergency model}	
%	\resizebox{0.95\linewidth}{!}{
	\centering
		\begin{tabular}{p{1,5cm}p{2 cm}p{4,0cm}}
	\doubleheavyrule
		\textbf{Opcodes}			& \textbf{Messages} 		&  \textbf{ Description}	\\
		\hline
		0xE1		& 0x0A					& \textit{Message to request help}	\\
		0xE2		& 0x0B				& \textit{Message to offer help}	\\
		0xE3			&  0x0C & \textit{Message to send a user status}	\\
		\doublerule
	\end{tabular}
	\label{tab:vendormodel}%
%	\vspace{-0.1cm}
%	}
\end{table}
\begin{figure}[!b]
	\centering
	\includegraphics[width=0.5 \textwidth]{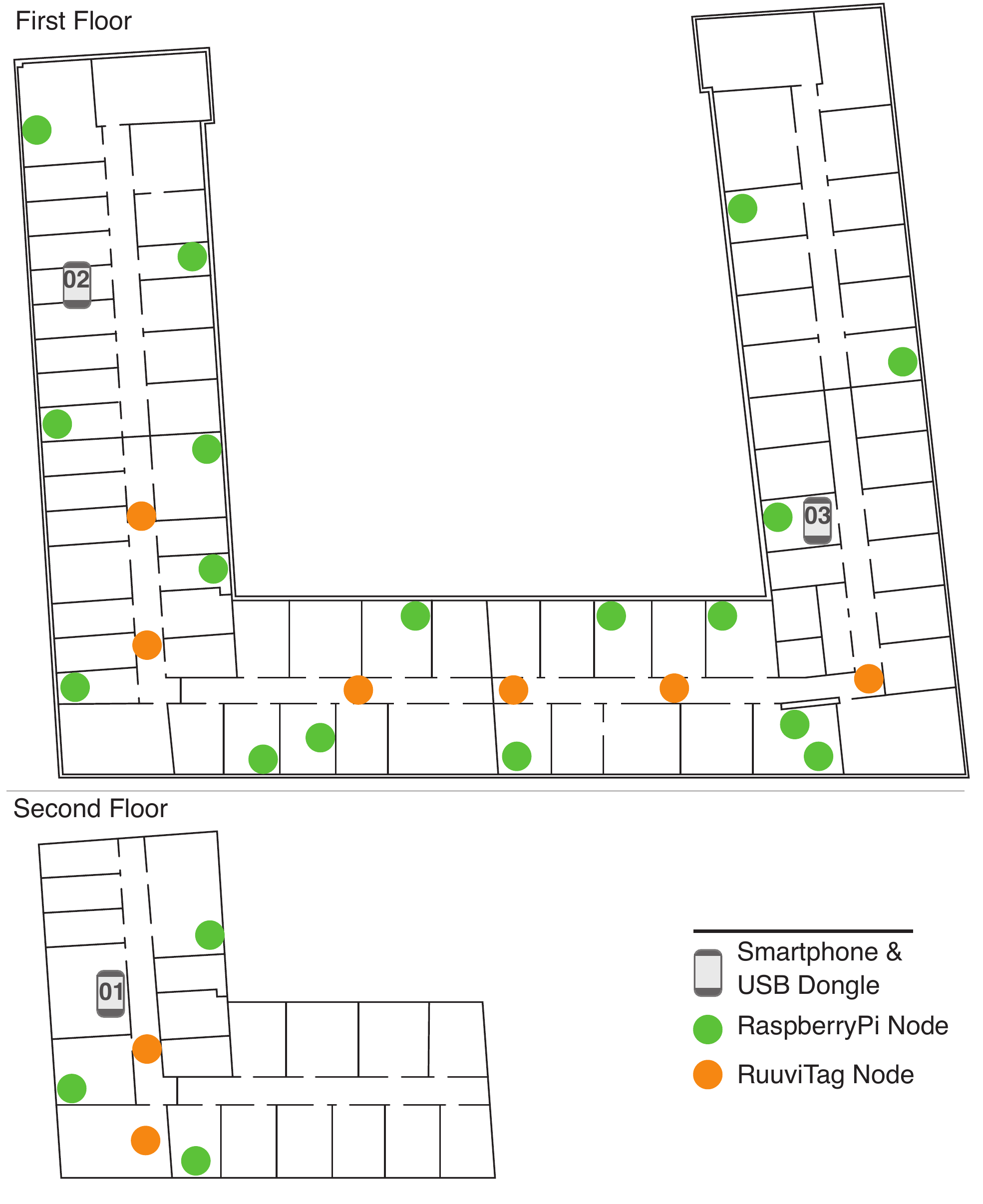}
	\caption{Proof-of-concept setup for the smart office experiments.}
	\label{testbedSO}
\end{figure}

Because all mesh packets are encrypted, a node without the security credentials can neither join the mesh network nor send/receive data to other nodes. 
To address this, we integrate a QR-Code reader interface to get the minimum required security credentials to join the network.
The QR-Code consists of a JSON format data that stores the security credentials needed to be part of the \gls{btmesh} network.
These credentials include: the network key, application keys for the vendor model, and an index which is needed to identify 
the subnetwork.

%!TEX root = ../ghtc_btmesh_2019.tex
\section{Proof-Of-Concept}
\label{system}
In this section we describe in detail our proof-of-concept implementation, as well as the hardware and software utilized.
Fig.~\ref{screenshots} illustrates our Android application developed to test the feasibility of \gls{bemergency}. We validate the communication between smartphones devices using a \gls{btmesh} network from two smart scenarios: \textbf{\textit{(A)}} - smart office, and \textbf{\textit{(B)}} - smart home.
\subsection{Hardware Setup}
The testbed consists of RuuviTags \cite{RuuviTag} sensors based on the nRF52832 SoC from Nordic Semiconductor, Nordic Semiconductor nRF52840 USB Dongles \cite{usbdongle}, Raspberry Pi 3 Model B+ \cite{raspberry} nodes based on the Broadcom BCM2837B0 SoC and smartphones Nexus 6P running Android version 8.1.0. 
\begin{table}[htbp!]
\renewcommand{\arraystretch}{1.5}
	\caption{Node features configured in the testbed}	
	\label{tab:capnodes}%
%	\resizebox{0.5\textwidth}{!}{
	%\centering
	\begin{tabular}{p{2,5cm}ccc}
			\doubleheavyrule
						& \textbf{RuuviTags/} 	&  \textbf{Linux Pis*}	& \textbf{Smartphones}\\
						& \textbf{USB Dongles} 	& 	& \\\hline
		Relay			& \CIRCLE							& \CIRCLE				& \Circle\\
		Proxy			& \CIRCLE							& \Circle				& \Circle\\
		GATT Bearer		& \CIRCLE							& \Circle				& \CIRCLE\\
		Adv. Bearer		& \CIRCLE							& \CIRCLE				& \Circle\\
		\doublerule		
		\\
		\multicolumn{4}{l}{
			\CIRCLE \: 		{fulfills feature}, 
%			\RIGHTcircle \: fulfills, but not used
			\Circle \: 		does not fulfill feature
		}
		\\
		\multicolumn{4}{l}{* Pis were used only in the smart office scenario}
	%	\vspace{-0.4cm}
	\end{tabular}
%	}
\end{table}

Table \ref{tab:capnodes} summarizes the node features configured on each device for both scenarios.
Because the Raspberry Pis support only the relay features, we use the proxy protocol with Nordic USB Dongles.

For simplicity, an additional smartphone is initially used as \textbf{\textit{provisioner}}.

\subsubsection{Smart office scenario}
Fig.~\ref{testbedSO} visualizes the location of the nodes on scenario \textbf{\textit{A}}.
The nodes are distributed throughout an office building over two adjacent floors, each floors consists of offices and meeting rooms. 
Due to the high density of WiFi access points as well as other equipment operating in the 2.4GHz band, the nodes have to cope with high interference.
In the first floor the nodes are arranged in an area of approximately 900 m$^{2}$, and in the 2nd floor the overall facility measures approximately 180 m$^{2}$. 
The maximal distance between two nodes is approx. 10 m and the minimal distance is close to 1 m. 

\subsubsection{Smart home scenario}
\begin{figure}
	\centering
	\includegraphics[width=0.5 \textwidth]{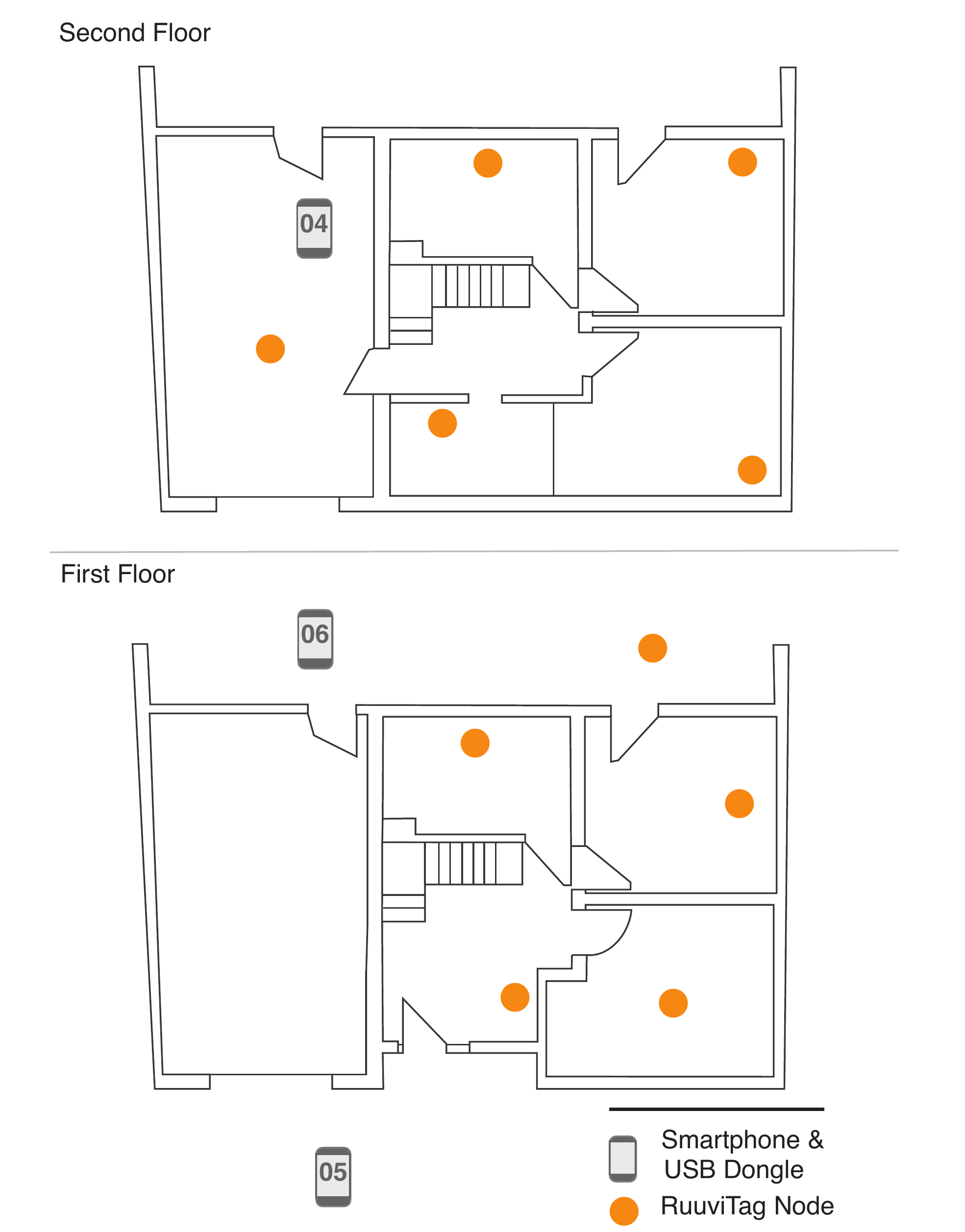}
	\caption{Proof-of-concept setup for the smart home experiments.}
	\label{testbedSH}
\end{figure}

Fig.~\ref{testbedSH} shows the proof-of-concept setup for scenario \textbf{\textit{B}}. We distribute the nodes in a brick house with two floors in a residential area.
The area covered by the smart home installation is approximately 63 m$^{2}$ per floor. 
The maximal distance between two nodes is approx. 6.5 m and the minimal distance is approx. 3 m.

\subsection{Software }
For our experiments, we use the SDK Softdevice version 6.1.0\cite{nordicsdk} and the Mesh SDK version 3.1.0 \cite{nordicbtmeshsdk}, both developed by Nordic Semiconductor. 
The Android-nRF-Mesh library \cite{androidmesh} is utilized for the initial setup configuration (provisioning phase). 
We build and extend the RuuviTag firmware from the Git repository \cite{ruuviblog} to integrate the mesh stack.
For supporting mesh on the Raspberry Pis, BlueZ\cite{BlueZ} version 5.50 was extended and rebuilt.
Additionally, we integrate the Nordic library \cite{androidmesh} to our smartphone application to support the proxy protocol as well as the proxy client on the smartphones.
\subsection{Support for the proxy protocol}
Currently, the Android Bluetooth stack does not provide the \gls{btmesh} stack neither the proxy protocol built in. 
To address this, we integrate the Android-nRF-Mesh library \cite{androidmesh} developed by Nordic Semiconductor into our smartphone application. 
The nRF-Mesh library supports the proxy protocol on Android devices only for the network configuration phase.
In order to enable mobile devices to participate in an existing \gls{btmesh} network, we implement and integrate the 
proxy functionality specified in the standard into our Android application.
With these changes, a smartphone can receive and deal with \gls{btmesh} messages. 

\subsection{Network configuration phase}
As mentioned before, a \textbf{\textit{provisioner}} is responsible for the initial setup and any reconfiguration of the nodes in the network. For the experiments, we consider an already existing \gls{btmesh} network, both in the smart home as well as in the smart office scenario.
We also implement a scanning QR-Code functionality, to allow smartphone devices to be part of the existing network by only scanning the required security materials. 

\subsection{Network services}
For the experiments, we consider the following configuration: each RuuviTag and Raspberry Pi implement and enable the relay feature.
The USB Dongles act as proxy server, i.e., they implement and enable the proxy feature.
Thus the smartphones communicate with the USB Dongles to send/receive mesh messages.
Because the Android application implements our vendor model, the smartphones can exchange messages between them using the existing \gls{btmesh} network. 
For simplicity, we set the destination address to predefined broadcast address.
So each node that receives a message and implements our model can process it. 

%!TEX root = ../ghtc_btmesh_2019.tex
\section{Evaluation}
\label{evaluation}
In this section, we show the feasibility of our solution that leverages smart environments to help forming post-disaster communication networks.
To this end, we implement a proof-of-concept application and test it in combination with the two outlined \gls{btmesh} scenarios using real devices.

\subsection{Procedure}
We perform a set of experiments in order to evaluate the performance of a \gls{btmesh} network regarding
 packet loss and response time. Each interaction from the experiments implies a variation of the number of messages sent: 
\textit{experiment I:} first, we send 5 messages per minute, \textit{experiment II:} we increase the number of 
messages to 10 messages per minute, 
and finally, \textit{experiment III:} we send 20 messages per minute. Each experiment runs for 12 minutes.
We repeat this procedure 5 times. Detailed experiment settings are provided in Table \ref{tab:experiments}.

\begin{table}
\renewcommand{\arraystretch}{1.5}
\smallskip
\smallskip
	\caption{Proof-of-concept settings}	
	\label{tab:experiments}
%	\resizebox{0.99\linewidth}{!}{
	\centering
	\begin{tabular}{p{1,5cm}p{3,5cm}p{2,5cm}}
	\doubleheavyrule
		\textbf{Scenario A}		& Dimensions w x h					& 13.6 x 9.25 [m]\\
								& Number of relay nodes				& 8\\
								& Distance between nodes (max, min)	& (6.5, 3) [m]\\\hline
		\textbf{Scenario B}		& Dimensions w x h					& 85 x 65 [m]\\
								& Number of relay nodes				& 28\\
								& Distance between nodes (max, min)	& (10, 1) [m]\\\hline
		\textbf{Both}	& Number of proxy servers			& 3\\
								& Number of proxy clients			& 3\\
								& Models							& emergency model\\
								& Messages sent per minute	& 5 - Experiment I\\
								&		 							& 10 - Experiment II\\
								&									& 20 - Experiment III\\
		\doublerule
	\end{tabular}
\end{table}
\subsubsection{Smart office scenario}
We first configure \textbf{\textit{01}} as the source node which generates the \gls{btmesh} messages. 
It sends a help request message to all nodes in the network, in our case to the other smartphones. 
As illustrated in Fig.~\ref{testbedSO}, \textbf{\textit{01}} is located on the second floor and the other nodes \textbf{\textit{02, 03}} are located on the first floor. 
These nodes respond to the help request by confirming that they offer help. 

\subsubsection{Smart home scenario}
In addition to the smart office scenario, a smart home experiment was carried out. 
As depicted in Fig.~\ref{testbedSH}, node \textbf{\textit{04}} was located inside the house, and nodes \textbf{\textit{05, 06}} were located outside the house in close proximity. 
As a result, the smartphone outside the house were able to connect with the \gls{btmesh} network and to reach any device locate inside the house.
%Allowing the smartphone to get a connection with the \gls{btmesh} network and so be able to send messages to any another device located inside the house.

\subsection{Results}
Table \ref{tab:results} summarizes the most important results from our experiments. 
\begin{table}[!t]
\renewcommand{\arraystretch}{1.5}
	\caption{Experiment results}	
	\label{tab:results}
%	\resizebox{0.99\linewidth}{!}{
	\centering
	\begin{tabular}{p{3cm}ccc}	
	\doubleheavyrule
	\textbf{Smart office} & & & \\\hline													
	 \textbf{Metric}		& \textbf{Mean} & \textbf{Standard deviation}	& \textbf{Median}\\\hline
		Number of hops		& 6.15			& 1.43							& 6.0\\
							 Response time	[ms]	& 1053.13		& 453.20						& 1020.0\\
							 Packet loss rate	& 38.21		& 17.75						& 35.4\\
	\doubleheavyrule
	
	\textbf{Smart home} & & & \\\hline
	 \textbf{Metric}		& \textbf{Mean} & \textbf{Standard deviation}	& \textbf{Median}\\\hline
		 Number of hops		& 3.11			& 0.32							& 3.0\\
							 Response time	[ms]	& 995.53		& 349.60						& 827.5\\
							 Packet loss rate & 8.5		& 4.67						& 11.2 \\

		\doublerule
	\end{tabular}
\end{table}

The main goal of the experiments was to measure the response time to a help request as well as the packet loss rate in a real-world environment, including external interference, i.e., \gls{ble} devices such as another smartphones, WiFi devices, etc.

\begin{figure}[!t]
	\centering
	\includegraphics[width=0.5 \textwidth]{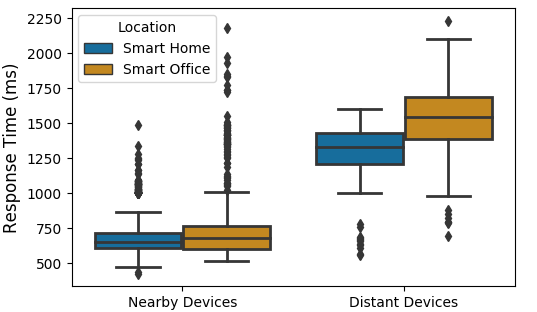}
	\caption{Response Time to a help message in both scenarios.}
	\label{resptime}
\end{figure}

Fig.~\ref{resptime} visualizes the response time to a help request in both smart environments. 
We can observe that the response time is directly influenced by the location of the nodes. 
As the distance between the nodes increases, the response time also grows. This is expected, as 
a message needs to traverse more hops to reach the destination.
Furthermore, each node that relays a message implies additional processing time.
The response time is in the order of one second for devices in proximity and increases to around 1.5 seconds for distant devices.
While these latencies are considerably higher than latencies in infrastructure networks, we consider them to be acceptable in post disaster scenarios, where the fact that communication and basic services are available at all can be considered paramount to minimizing latency.
%	\vspace{-0.1cm}

%\vspace{-0.1cm}
\begin{figure}[!t]
	\centering
	\includegraphics[width=0.5 \textwidth]{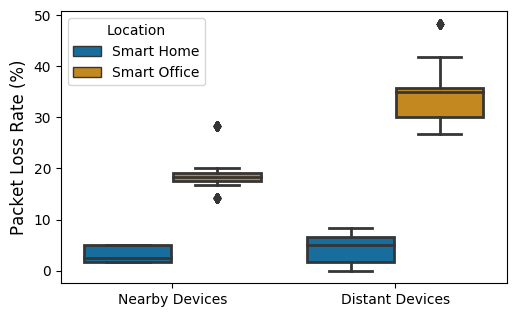}
	\caption{Packet loss rate in both scenarios.}
	\label{pktloss}
\end{figure}

Fig.~\ref{pktloss} shows the percentage of packet loss for each experiment. 
Although the packet loss rate for the smart home scenario indicates a similar pattern, it differs in the smart office scenario. 
On the one hand, the packet loss rate in the smart home scenario is almost constant. This is expected, as the density of other equipment operating in the 2.4 GHz band is very low.
On the other hand, we can notice that the distance between the nodes also impact the packet loss rate in the smart office scenario.
This result is reasonable, as during work hours there are a lot of additional \gls{ble} and WiFi devices such as notebooks, smartwatches, etc., that generate interfering transmissions in the 2.4 GHz band.

%!TEX root = ../ghtc_btmesh_2019.tex
\section{Discussion and Conclusion}
\label{conclusion}

In this paper, we showed that smart environments as found in today's and future digital cities can contribute in establishing post-disaster networks.
In particular, we showed that the novel \gls{btmesh} standard, which is supported by a wide range of \gls{iot} solutions, can be used to mediate post-disaster device-to-device communication even using most of today's smartphones.
We demonstrate the feasibility of such a system on common off-the-shelf devices, by designing and implementing our \gls{bemergency} proof-of-concept system.
To this end, an Android application implements the proxy protocol specified in the standard. 
Additionally, we propose an emergency model to enable smartphones exchange data using existing \gls{iot} devices.

We show the feasibility and performance of our solution in two \gls{btmesh} realistic scenarios, namely a smart office and a smart home scenario.
For the performance evaluation, we utilized heterogeneous \gls{iot} devices, i.e., Linux-based devices and novel devices that integrate the \gls{btmesh} stack directly in the firmware, together with regular Android smartphones that do not offer native \gls{btmesh} support.

By utilizing \gls{btmesh} as mediating technology, we can address the lack of direct communication between nearby mobile devices without the need to modify such devices, e.g., for supporting Wi-Fi in ad-hoc mode the devices must be jail-breaking. 
Finally, our experiments facilitate a first performance analysis of such a system.

While our experiments show the feasibility of the proposed \gls{bemergency} concept, we envision a number of improvements in future work.
For instance, the proposed emergency services could be enriched by location information to help discovering persons in need.
Since \gls{btmesh} has never been designed for emergency use, a number of other challenges remain. 
As surveyed in \cite{alvarezghtc16}, security does not lose importance during disasters. While security is a mandatory \gls{btmesh} feature, i.e., without the corresponding security credentials a device can neither join a mesh network nor exchange data with other nodes, it still lacks on usability during emergency situations.
For practical applicability, easy to use device-to-device security solutions could be integrated into our \gls{bemergency} concept, e.g., as proposed in \cite{alvarez2018sea}.

\section*{Acknowledgment}
This work has been supported by the German Federal Ministry of Education and Research (BMBF) and the Hessen State Ministry for Higher Education, Research and the Arts (HMWK) under the joint grant for the National Research Center for Applied Cybersecurity. It has further been supported by the German Research Foundation (DFG) as part of the project A3 within the Collaborative Research Center (CRC) 1053 – MAKI.

\bibliographystyle{IEEEtran}
\bibliography{IEEEabrv,ghtc2019}

\end{document}